\documentclass{pasj01}

\usepackage{color}

\input{pasjsup.sty}

\def\commenta{$^*$}
\def\commentb{$^\dagger$}
\def\commentc{$^\ddagger$}
\def\commentd{$^\S$}
\def\commente{$^\|$}
\def\commentf{$^\#$}
\def\commentg{$^{**}$}
\def\commenth{$^{\dagger\dagger}$}

\newcounter{author}
\def\authorcount#1#2{\refstepcounter{author}\label{#1}
                     \altaffiltext{\ref{#1}}{#2}}

\begin{document}
\SetRunningHead{K. Isogai et al.}{ASASSN-14dx}

\Received{201X/XX/XX}
\Accepted{201X/XX/XX}

\title{Third Nearest WZ Sge-Type Dwarf Nova candidate ASASSN-14dx Classified on the Basis of Gaia Data Release 2
}
\author{
  Keisuke~\textsc{Isogai},\altaffilmark{\ref{affil:Kyoto}*}
  Taichi~\textsc{Kato},\altaffilmark{\ref{affil:Kyoto}}
  Akira~\textsc{Imada},\altaffilmark{\ref{affil:Imada}}$^,$\altaffilmark{\ref{affil:HidaKwasan}}
  Tomohito~\textsc{Ohshima},\altaffilmark{\ref{affil:Nishiha}}
  Naoto~\textsc{Kojiguchi},\altaffilmark{\ref{affil:Kyoto}}
  Ryuhei~\textsc{Ohnishi},\altaffilmark{\ref{affil:Kyoto}}
  Franz-Josef~\textsc{Hambsch},\altaffilmark{\ref{affil:GEOS}}$^,$\altaffilmark{\ref{affil:BAV}}$^,$\altaffilmark{\ref{affil:Hambsch}}
  Berto~\textsc{Monard},\altaffilmark{\ref{affil:Monard}}
  Seiichiro~\textsc{Kiyota},\altaffilmark{\ref{affil:Kis}}
  Hideo~\textsc{Nishimura},\altaffilmark{\ref{affil:Nmh}}
  Daisaku~\textsc{Nogami}\altaffilmark{\ref{affil:Kyoto}}
}

\authorcount{affil:Kyoto}{
Department of Astronomy, Kyoto University, Kyoto 606-8502}
\email{$^*$isogai@kusastro.kyoto-u.ac.jp}

\authorcount{affil:Imada}{
  Hamburger Sternwarte, Universit\"{a}t Hamburg, Gojenbergsweg 112, D-21029 Hamburg,Germany}

\authorcount{affil:HidaKwasan}{
     Kwasan and Hida Observatories, Kyoto University, Yamashina,
     Kyoto 607-8471}

\authorcount{affil:Nishiha}{
        Nishi-Harima Astronomical Observatory, University of Hyogo, Japan}

\authorcount{affil:GEOS}{
     Groupe Europ\'een d'Observations Stellaires (GEOS),
     23 Parc de Levesville, 28300 Bailleau l'Ev\^eque, France}

\authorcount{affil:BAV}{
     Bundesdeutsche Arbeitsgemeinschaft f\"ur Ver\"anderliche Sterne
     (BAV), Munsterdamm 90, 12169 Berlin, Germany}

\authorcount{affil:Hambsch}{
     Vereniging Voor Sterrenkunde (VVS), Oude Bleken 12, 2400 Mol, Belgium}

\authorcount{affil:Monard}{
     Bronberg and Kleinkaroo Observatories, Center for Backyard Astrophysics Kleinkaroo,
	Sint Helena 1B, PO Box 281, Calitzdorp 6660, South Africa }

\authorcount{affil:Kis}{
     VSOLJ, 7-1 Kitahatsutomi, Kamagaya, Chiba 273-0126, Japan}

\authorcount{affil:Nmh}{
     Miyawaki 302-6, Kakegawa, Shizuoka 436-0086, Japan}




\maketitle

\begin{abstract}
ASASSN-14dx showed an extraordinary outburst whose features are 
the small outburst amplitude ($\sim$ 2.3 mag) and long duration ($>$ 4 years).
Because we found a long observational gap of 123 d before the outburst detection,
we propose that the main outburst plateau was missed and that
this outburst is just a ``fading tail'' often seen after the WZ Sge-type superoutbursts.
In order to distinguish between WZ Sge and SU UMa-type dwarf novae (DNe),
we investigated {\it Gaia} DR2 statistically.
We applied a logistic regression model and succeeded in classifying
by using absolute {\it Gaia} magnitudes $M_{G}$ and {\it Gaia} colors $G_{\rm BP}-G_{\rm RP}$.
Our new classifier also suggests that ASASSN-14dx is the best candidate of a WZ Sge-type DN.
We estimated distances from the Earth of known WZ Sge stars by using {\it Gaia} DR2 parallaxes.
The result indicates that ASASSN-14dx is the third nearest WZ Sge star (next to WZ Sge and V455 And),
and hence the object can show the third brightest WZ Sge-type superoutburst whose maximum is $V$ = 8--9 mag.
\end{abstract}

\section{Introduction}\label{sec:intro}
Cataclysmic variables (CVs) are close binaries composed of 
a white dwarf (WD) primary and a Roche lobe-filling secondary.
An accretion disk is formed around the primary
because the secondary transfers its mass to the primary.
Dwarf Novae (DNe), which are a subclass of CVs, are characterized by
a sudden increase of the disk brightness called ``outburst''.
There are two competing models to explain a dwarf nova outburst:
the mass-transfer burst model \citep{bat73DNmodel}
and the thermal instability model \citep{osa74DNmodel,hos79DImodel}.
At the present time, the latter is widely accepted
at least for typical dwarf nova outbursts.
For instance, \citet{dub18gaia_arxiv} estimated the mass-transfer rates
of 130 CVs by using the parallax distances
of the {\it Gaia} Data Release 2 (DR2).
The estimated mass-transfer rates of dwarf novae and
nova-like stars are consistent with their theoretical values
predicted by the thermal instability model, respectively.
(for a review of CVs and DNe, see e.g. \cite{war95book}).

SU UMa-type DNe show not only normal outbursts
but also superoutbursts with larger scales.
During superoutbursts, we can observe ``superhumps'' whose amplitude is 0.1--0.5 mag
and period is a few percent longer than the orbital period $P_{\rm orb}$.
Superoutbursts and superhumps are thought to be a result of the thermal-tidal instability \citep{osa89suuma}.
When an outburst begins, the disk expands.
If the object has a sufficiently low mass ratio $q$,
the outer edge of the disk can reach the 3:1 resonance radius
which causes the tidal instability \citep{whi88tidal,osa89suuma,lub91SHa,lub91SHb,hir90SHexcess}.

WZ Sge-type DNe are a subclass of SU UMa-type ones
and considered to be objects at the terminal stage of the CV evolution.
WZ Sge stars are characterized by the large amplitude and long duration superoutbursts which
are accompanied by ``early superhumps'' in the early terms of the superoutbursts.
According to the modern criteria provided by \citet{kat15wzsge},
WZ Sge stars are defined by the presence of an early superhump phase.
It is thought that early superhumps are triggered
by the 2:1 resonance \citep{lin79lowqdisk,osa02wzsgehump}.
Early superhumps are double-peaked modulations whose periods are close to $P_{\rm orb}$.
Because the early superhumps are caused by a geometrical effect,
their amplitudes depend on their inclination angles.
Thus, in low-inclination systems, we can observe a long outburst plateau
with no superhumps instead of an early superhump phase.
SU UMa stars evolve into WZ Sge ones as the mass transfer proceeds
because only CVs with extremely low mass ratios (typically $q < 0.09$)
and low mass-transfer rates can reach the 2:1 resonance radius,
(for a review of WZ Sge-type DNe, see \cite{kat15wzsge}).

According to the theoretical CV evolution calculated by \citet{kol93CVpopulation},
70$\%$ of CVs have already passed the period minimum. 
Such objects are called ``period bouncers''.
The secondary masses of the period bouncers are below the substellar limit.
Their $q$ and mass-transfer rates are thus extremely low
as compared with DNe before the period minimum.
Empirically, we know many objects around the period minimum show WZ Sge-type superoutbursts.
These facts imply that most of DNe are WZ Sge stars.
However, the number of WZ Sge stars is much smaller than that of SU UMa ones.
The gap between the theory and observation can be caused by the observational bias
because WZ Sge stars are fainter and their outburst intervals are longer in comparison to SU UMa ones.
Thus, it is important to dig up the hidden WZ Sge stars.
In addition, there are some unsolved problems in WZ Sge stars,
e.g. the detailed mechanism of early superhumps or rebrightenings.
The increasing number of WZ Sge stars will give us more opportunities to observe such phenomena.

We report on the outburst and classification of ASASSN-14dx.
The outburst was detected at $V=$ 13.95 on June 25, 2014 by the All-Sky Automated Survey for Supernovae (ASAS-SN) \citep{ASASSN}.
The coordinates are RA = 02:34:27.73 and Dec = -04:54:30.7 at J2000.0.
The Sloan Digital Sky Survey (SDSS) magnitudes are $u=16.42, g=16.26, r=16.37, i=16.55 z=16.63$ \citep{SDSS9}.
The object is an XMM-Newton source and has a GALEX counterpart
with an NUV and FUV magnitudes of 16.4 and 16.6 \citep{GALEX}.
The spectra were obtained in August 2014 \citep{14dxAtel} and in October 2014 \citep{tho16CVs}.
They reported double-peaked emission lines with broad and narrow absorption ones
on a blue continuum, and thus the object was confirmed as a CV.
Since follow-up photometries by our group and \citet{har1714dx}
reported the superhump-like modulations,
the object was initially classified as an SU UMa star.
\citet{har1714dx} confirmed that the object doesn't show any eclipses.
On the basis of our light curve analyses and new classification method
by using the {\it Gaia} Data Release 2 (DR2) \citep{Gaia2016,Gaia2018},
we propose that ASASSN-14dx is a WZ Sge star rather than SU UMa one.

\section{Observation and Analysis}\label{sec:obs}
Our time-series observations are summarized in table E1.
The data were acquired by time-resolved unfiltered CCD photometry 
using 30-40cm class telescopes by the VSNET Collaboration \citep{VSNET}.
The times of the observations were corrected to Barycentric Julian Date (BJD).
We adjusted the zero-point of each observer to Franz-Josef Hambsch's data.

We used the phase dispersion minimization (PDM) method 
for analyzing the periodic modulations.
We estimated $1\sigma$ errors by using the methods
in \citet{fer89error} and \citet{Pdot2}.
Before our period analyses, we subtracted the global trend of the light curve
which was calculated by using locally-weighted polynomial regression (LOWESS, \cite{LOWESS}).

\begin{figure}
\begin{center}
    \FigureFile(80mm,50mm){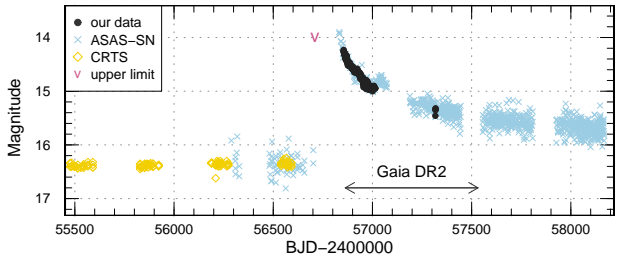}
\end{center}
\caption{Overall light curves of ASASSN-14dx.
Filled circles, crosses, open diamonds and ``V''-mark represent our observations,
the ASAS-SN data, the CRTS data and the upper limit, respectively.
Our time-series observations were binned to 0.01 d.
The observation period of {\it Gaia} DR2 is indicated by the arrows.}
\label{fig:lc}
\end{figure}

\section{Results}\label{sec:result}
\subsection{Overall Light Curve}\label{sec:outburst}
Figure \ref{fig:lc} shows the overall light curve of ASASSN-14dx.
The light curve also includes the public data of the CRTS \citep{CRTS}
and ASAS-SN Sky Patrol \citep{ASASSN,koc17ASASSNLC}.
Before the outburst, the object was stable around $V=16.36$ mag.
The outburst was detected at $V=13.95$ mag on June 25, 2014 (BJD 2456833),
then the object gradually decreased to $V \sim 15.7$ mag until 2018.
Such a small-amplitude and long-duration outburst is very unusual for DN outbursts.

We found an upper limit of 14.0 mag on February 22, 2014 (BJD 2456710) 
indicated in figure \ref{fig:lc} as ``V''-mark.
After this observation, we don't have any data until the outburst detection due to seasonal reason,
namely, there is a long observational gap of 123 d.
This gap implies that we missed the initial phase of the outburst
and that this outburst is just a ``fading tail''.

A fading tail is a well-known phenomenon seen after WZ Sge-type superoutbursts.
Specifically, \citet{kuu11wzsge} reported the moderate declines of WZ Sagittae
after the end of the superoutbursts plateau.
The brightness just after the superoutbursts is about 0.5 mag brighter than in quiescence.
They proposed that the decline can be caused by the development of a cavity
in the inner disk and that the time scale is about a decade.
For another example, SSS J122221.7-311523 (hereafter J1222) showed a fading tail lasting
at least 500 d and the brightness just after the superoutburst is about 3 mag brighter \citep{neu17j1222}.
The outburst profile of ASASSN-14dx is in good agreement with them rather than ordinary outbursts.

\begin{figure}
\begin{center}
    \FigureFile(80mm,50mm){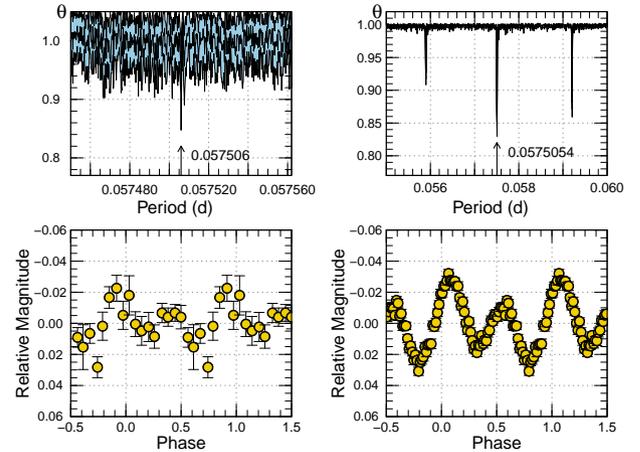}
\end{center}
\caption{Results of the period analyses of orbital modulations 
in quiescence (BJD 2453627.89--2456591.85) (Left) 
and outburst (BJD 2456854.82--2457014.09) (Right).
(Upper): $\theta$ diagram of our PDM analysis of superhumps.
The shaded area means $1 \sigma$ errors.
(Lower): Phase-averaged profiles of orbital modulations.
The zero phases were defined to be BJD 2456945.8837,
which is the time of the inferior conjunction obtained by \citet{tho16CVs}.}
\label{fig:pdm}
\end{figure}

\subsection{Double-wave modulations}\label{sec:superhump}
During the outburst, the object showed persistent double-wave modulations.
The result of the period analysis of our observations 
in BJD 2456854.82--2457014.09 is indicated in figure \ref{fig:pdm}.
The estimated period is 0.0575054(4) d.
The interval of two peaks is nearly 0.5 phase.
Since the modulations were initially interpreted as superhumps,
the object was classified as an SU UMa-type DN (cf. vsnet-alert 17598, \cite{har1714dx}).

\section{Discussion}\label{sec:discuss}
\subsection{Orbital Modulations}\label{sec:orbital}
The left panel of figure \ref{fig:pdm} shows the period analysis
of the CRTS data in quiescence (BJD 2453627.89--2456591.85).
We reported a weak signal of a possible orbital period of 0.0575060(2) d (vsnet-alert 18017).
\citet{tho16CVs} estimated the orbital period to be 0.05756(6) d from H$\alpha$ radial velocities
via spectroscopy, and they confirmed that this value is consistent with our result.
These values are close to the period detected during the outburst of 0.0575054(4) d. 
The averaged profiles are also similar to each other (the lower panels in figure \ref{fig:pdm}).
Thus we propose that the  modulations in the outburst
are not superhumps but orbital modulations.
We adopted the period in quiescence of 0.0575060(2) d as the orbital one
because of the small error and the long baseline of the data.

WZ Sge-type DNe, especially in high-inclination systems, often show
double-wave orbital modulations in quiescence or fading tail (e.g. \cite{pat96alcom}).
\citet{ski00wzsge} investigated the disk structure of WZ Sagittae
in quiescence from the Doppler mapping.
They found that the bright spot exists along the ballistic trajectory,
and considered that a low-density disk allows accretion stream to penetrate
and thus can explain double-wave modulations originated from the change of the line-of-sight.
\citet{kon15wzsgequihump} simulated the humps of WZ Sge stars
by using the three-dimensional computation.
They proposed that a spiral density wave and four kinds of shocks can cause such modulations.
Because the density wave undergoes a slow retrograde precession in the observer's frame
and the interaction between the density wave and four shocks generates the modulations,
they succeeded in reproducing the profile variations and phase shifts of humps, which are sometimes seen in WZ Sge stars.
Note that the period of such humps can be shorter than the orbital one in a strict sense.
The modulations of ASASSN-14dx could also be explained by these mechanisms.

The estimated period of ASASSN-14dx in the outburst is very slightly shorter than that in quiescence.
According to the thermal instability model, a disk after an outburst
gradually shrinks due to the accreting mass (cf. \cite{osa89suuma}).
Shrinkage of a disk could lead to a phase shift of the modulations, e.g. a shift of the position of the bright spot.
In principle, we cannot discriminate between a continuous phase shift and a period variation.
On the other hand, the model of \citet{kon15wzsgequihump}
can account for the shorter period owing to the change of the precession rate of the density wave.
The small difference in the periods might be caused by these varying disk structure.

We drew the averaged profiles of orbital modulations (the lower panels of figure \ref{fig:pdm}).
We estimated the time of the inferior conjunction,
the secondary star is the closest to the Earth, to be BJD 2456945.8837(7)
on the basis of the time-resolved spectroscopies obtained by \citet{tho16CVs},
and we defined the zero phases to be BJD 2456945.8837.
For the purpose of understanding the mechanism of the humps,
it may be worth mentioning that both of the peaks of the modulations are close to the conjunction.
However, we must note that the time of the conjunction is estimated by using the disk emissions,
whose light source is unclear.
Therefore this value has an uncertainty, and we need to observe
lines from the secondary star to obtain the accurate time of the conjunction.
These figures suggest that the phases of the peaks in quiescence and outburst
are respectively $-0.06(2)$ and $0.08(1)$, namely the object showed a phase shift of 0.14(2).
Note that the folding periods are different from each other.
However, even if we fold the light curve in outburst by using the period in quiescence,
the phase is little varied (less than 0.01 phase).
It may be difficult to explain the large shift only by a simple bright spot and shrinking disk,
while the model of \citet{kon15wzsgequihump} can explain it
on the basis of the retrograde-precession density wave.

\begin{figure}
\begin{center}
    \FigureFile(80mm,110mm){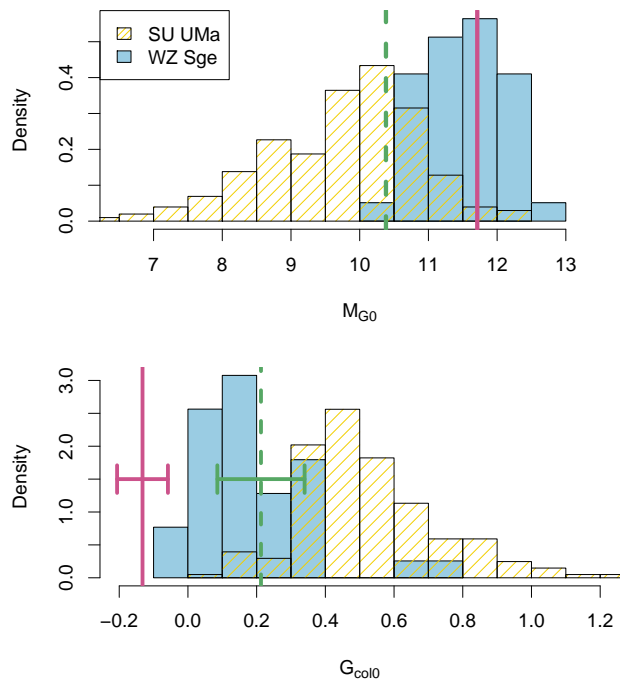}
\end{center}
\caption{Density distributions of corrected absolute {\it Gaia} magnitudes $M_{G0}$ (upper)
 and {\it Gaia} colors $G_{\rm col0}$ (lower).
Hatched and shaded boxes represent SU UMa and WZ Sge-type DNe, respectively.
Solid and dashed vertical lines indicate ASASSN-14dx in the SDSS (quiescence) and {\it Gaia} DR2 (outburst), respectively.
The error bars of ASASSN-14dx are shown only in the lower panel because of the large errors.
We removed three WZ Sge-type DN candidates mentioned in section \ref{sec:wzsgecandidate}.
}
\label{fig:gaia1}
\end{figure}

\begin{figure}
\begin{center}
    \FigureFile(80mm,100mm){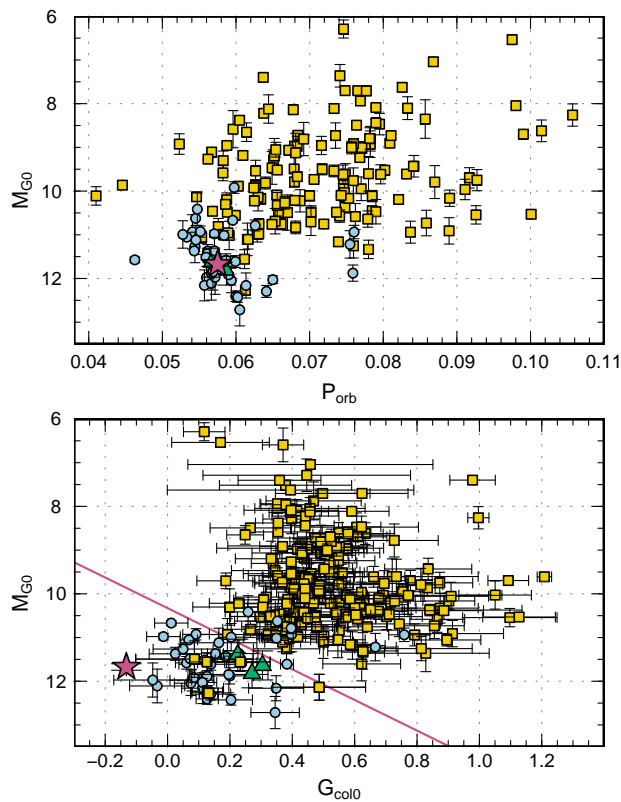}
\end{center}
\caption{Corrected absolute {\it Gaia} magnitude $M_{G0}$ versus orbital period $P_{\rm orb}$ (upper) and
{\it Gaia} color $G_{\rm col0}$ (lower).
Regarding WZ Sge-type DNe whose orbital periods are unknown,
we estimated the orbital periods from the superhump periods and equation (6) in \citet{Pdot3}.
The filled squares, circles and star represent SU UMa, WZ Sge-type DNe and ASASSN-14dx, respectively.
Three WZ Sge star candidates mentioned in section \ref{sec:wzsgecandidate} are shown as filled triangles.
The solid line in the lower panel indicates $P_{\rm WZ}=0.5$ which means the borderline between SU UMa and WZ Sge stars.}
\label{fig:gaia2}
\end{figure}

\subsection{Classification by {\it Gaia} DR2}\label{sec:classify}
In order to distinguish WZ Sge stars from SU UMa ones,
we investigated the known DNe of the {\it Gaia} DR2.
By using the parallaxes $\varpi$ and three broad band optical magnitudes
(white light $G$, blue $G_{\rm BP}$ and red $G_{\rm RP}$) provided by the {\it Gaia},
we can estimate the absolute {\it Gaia} magnitudes
 $M_{G}$ and {\it Gaia} colors $G_{\rm col} = G_{\rm BP}-G_{\rm RP}$
\citep{GaiaDR2parallax,GaiaDR2photmetry,GaiaDR2photometry2}.
Although the {\it Gaia} team has already studied on the color-absolute
magnitude diagram about variable stars \citep{GaiaDR2colorAbmagVariables},
we concentrated more on the subclasses of DNe.
The CV evolutionary track computed by \citet{kni11CVdonor} shows that
absolute magnitudes become fainter and that the colors become bluer as CVs evolve.
Actually, we can see WZ Sge-type DNe lying on the bluer area
in the color-color diagram of the SDSS \citep{kat12DNSDSS}.
Furthermore, it is considered that $M_{G}$ depends also on the mass-transfer rates from secondaries,
i.e. the brightness of bright spots and disks (cf. \cite{dub18gaia_arxiv}). 

We identified known SU UMa/WZ Sge-type DNe with the {\it Gaia} DR2 sources.
The list of known DNe was taken from the AAVSO VSX\footnote{
  $<$https://www.aavso.org/vsx/$>$.
}.
We only used objects whose $\varpi$ errors are within $20\%$.
The galactic extinctions in $V$ and $B$ bands ($A_{V}$ and $A_{B}$)
for each object were obtained from the NED\footnote{
The NASA/IPAC Extragalactic Database (NED)
is operated by the Jet Propulsion Laboratory, California Institute of Technology,
under contract with the National Aeronautics and Space Administration.
} \citep{sch11extinction}.
Because the galactic extinctions are affected by the positions of the objects,
we estimated the effects by using equation 2 in \citet{ak08CVdist},
the distances $d$ calculated from $\varpi$ and
the assumed scale height for the interstellar dust $H = 100$ pc.
Then we converted $A_{V}$ and $A_{B}$ to
the extinctions in {\it Gaia} magnitudes ($A_{G}, A_{\rm BP}, A_{\rm RP}$)
on the basis of equation 1 and table 1 in \citet{bab18gaiaHR}.
Eventually, we used the corrected absolute magnitudes
$M_{\rm G0} =  M_{G} - A_{G}$,  de-reddened colors
$G_{\rm col0} = G_{\rm col} - (A_{\rm BP} - A_{\rm RP})$
and $P_{\rm orb}$ for the statistics.
We note that the effects of galactic extinction have large uncertainties
and are sometimes neglected (e.g. \cite{GaiaDR2colorAbmagVariables}).
Thus we removed the objects with large extinctions $A_{G} > 0.5$ in our statistics to suppress the uncertainty.

Figure \ref{fig:gaia1} shows the density distributions of $M_{G0}$ and $G_{\rm col0}$.
Hatched and shaded boxes represent SU UMa and WZ Sge stars, respectively.
As expected by previous studies, WZ Sge stars are basically fainter and bluer.
These features can give the interpretations that the mass-transfer rate is lower and
that the WD's contribution to the color is more important in WZ Sge stars.
The upper panel implies that there is a lower limit of $M_{G0} \sim 10.5$ as for WZ Sge stars.
This fact is consistent with the interpretation of early superhumps by the TTI model.
In other words, a low mass-transfer rate is needed to accumulate enough mass
at the onset of an outburst to trigger the 2:1 resonance \citep{osa95wzsge}.

ASASSN-14dx is indicated by dashed lines in figure \ref{fig:gaia1}.
Their positions are a little far from other WZ Sge-type objects.
The reason is that the $Gaia$ obtained the data
during the outburst as indicated in figure \ref{fig:lc}.
In order to estimate the $Gaia$ magnitudes in quiescence,
we converted the SDSS magnitudes, were observed before the outburst,
to {\it Gaia} ones by using table A.2 in \citet{GaiaDR2photometry2}.
Then, we obtained the quiescent values of $M_{G0}= 11.711(43)$ and $G_{\rm col0}=-0.112(74)$
indicated by solid lines in figure \ref{fig:gaia1}.
These figures give the impression that ASASSN-14dx is indeed a WZ Sge-type DN.

Figure \ref{fig:gaia2} shows the relations between
$M_{G0}$ and $P_{\rm orb}$ (upper), and $M_{G0}$ and $G_{\rm col0}$ (lower).
The filled squares, circles and star represent SU UMa, 
WZ Sge-type DNe and ASASSN-14dx in quiescence, respectively.
The triangles indicate new WZ Sge-type DN candidates discussed in section \ref{sec:wzsgecandidate}.
We removed them from our statistics.
Three WZ Sge-type stars around $P_{\rm orb} \sim 0.076$ d are
two long-$P_{\rm orb}$ WZ Sge stars (RZ Leo and ASASSN-16eg) and one period bouncer J1222.
\citet{wak17asassn16eg} interpreted the long-$P_{\rm orb}$ WZ Sge stars as borderline objects
between SU UMa and WZ Sge-type DNe, which have low mass-transfer rates in spite of their high-mass ratios.
They are indeed fainter than other SU UMa stars having similar $P_{\rm orb}$.
On the other hand, their colors are redder in comparison with the other WZ Sge stars
 (see two objects with $G_{\rm col0}$ of 0.6--0.8 in the lower panels of figure \ref{fig:gaia1} and \ref{fig:gaia2}).

From the lower panel of figure \ref{fig:gaia2},
we established a classification method between WZ Sge and SU UMa stars
by using a logistic regression model \citep{logistic,logistic2}.
On the basis of the data in the lower panel, we can estimate
the WZ Sge-type DN probability $P_{\rm WZ}$ as a function of $M_{G0}$ and $G_{\rm col0}$:
\begin{equation}\label{eq:1}
P_{\rm WZ} = \frac{1}{1 + \exp(19.8573 + 6.7335 G_{\rm col0} - 1.9227 M_{G0})}.
\end{equation}
The solid line in figure \ref{fig:gaia2} represents $P_{\rm WZ}=0.5$.
We propose $P_{\rm WZ}=0.5$ as the classifier of WZ Sge/SU UMa-type DNe.
For the sake of convenience, we obtained the WZ Sge probability without 
taking the galactic extinctions into account:
\begin{equation}\label{eq:2}
P'_{\rm WZ} = \frac{1}{1 + \exp(15.9452 + 7.5695 G_{\rm col} - 1.5448 M_{G})}.
\end{equation}
We note that this equation should not be used for high-$A_{G}$ objects.
We provide a sample R code to estimate $P_{\rm WZ}$ and the extinctions
on our website\footnote{http://www.kusastro.kyoto-u.ac.jp/\~{}isogai/Pwz/}.

We estimated $P_{\rm WZ}$ for each object including large-$A_{G}$ objects and summarized in table E2 and E3.
$P_{\rm WZ}$ of ASASSN-14dx is 0.21(14) in the Gaia DR2 (outburst)
but 0.97(1) in the SDSS (quiescence).
Our new classifier also suggests that ASASSN-14dx is the best candidate of a WZ Sge star.
We evaluated the validity of the $P_{\rm WZ}=0.5$ method, then
32 in 41 WZ Sge stars and 219 in 225 SU UMa ones are correctly classified,
whereas $P'_{\rm WZ}=0.5$ classified 30 WZ Sge ones and 220 SU UMa ones.
However, some WZ Sge stars could be classified as SU UMa ones due to the lacking data.
Intensive observations for high-$P_{\rm WZ}$ systems will improve our classification method.

\subsection{Exploration of WZ Sge-type DN candidates}\label{sec:wzsgecandidate}
We concluded that the following three objects are good candidates
of WZ Sge-type DNe on the basis of $P_{\rm WZ}$ and the modern criteria provided in \citet{kat15wzsge}:

\begin{itemize}
\item PNV J19321040-2052505\\
The estimated $P_{\rm WZ}$ is 0.63(18).
This object was detected on 2017 Nov. 11 by H. Nishimura\footnote{
  $<$http://www.cbat.eps.harvard.edu/unconf/followups/J19321040-2052505.html$>$.
}.
A follow-up observation confirmed a possible signal of superhumps,
and hence the object was classified as an SU UMa-type DN (vsnet-alert 21694).
According to ASAS-3 data, the object showed an outburst in 2003 \citep{ASAS3}.
The duration of the outburst was longer than 27 days and
the outburst amplitude was large ($\sim 8$mag).
These values strongly suggest that the object is a WZ Sge-type DN.

\item CRTS J200331.3-284941\\
The estimated $P_{\rm WZ}$ is 0.74(19).
The ordinary superhumps were detected in the 2015 outburst.
\citet{Pdot8} speculated that this object can be a WZ Sge-type DN
on the basis of the low mass ratio of 0.084(1), long duration of growing (stage A) superhumps
(more than 50 cycles), fading tail like WZ Sge-type DNe and long observational gap.

\item ASASSN-17kg\\
The estimated $P_{\rm WZ}$ is 0.62(17).
This object was detected on Jul. 31, 2016 by the ASAS-SN and
the last observation before the detection was on Jul. 27 \citep{ASASSN}.
The follow-up observation on Aug. 3 detected growing superhumps (vsnet-alert 21326).
In other words, there was an observational gap of seven days.
Although the typical duration of the early superhump stage is 10 d or more \citep{kat15wzsge},
table 9 in \citet{nak13j2112j2037}, which listed those of well-observed WZ Sge-type DNe,
shows the minimum duration is 5 d.
This fact suggests that the observational gap of one week could conceal the early superhump phase.
Moreover, the object showed one rebrightening after the superoutburst (vsnet-alert 21360).
We thus concluded that this object is a WZ Sge-type DN candidate.
\end{itemize}

Other SU UMa stars with a long observational gap could also be WZ Sge ones.
In future, we should pay attention to high-$P_{\rm WZ}$ objects
in order not to miss the early superhump phase.
Note that there are some unusual WZ Sge stars.
It may be difficult to discover them on the basis of $P_{\rm WZ}$.
Although AL Com was considered to be a typical WZ Sge star,
the object showed SU UMa-type superoutburst in 2015 \citep{kim16alcom}.
The estimated $P_{\rm WZ}$ of 0.11(19) is indeed low
(we removed AL Com from our statistics due to the large error of $\varpi$).
As mentioned in section \ref{sec:classify},
the long-$P_{\rm orb}$ WZ Sge stars (RZ Leo, ASASSN-16eg and BC UMa)
have large $G_{\rm col0}$, and hence the $P_{\rm WZ}$ are less than 0.15.
In addition, the {\it Gaia} DR2 data are possibly contaminated by outbursts,
eclipses and so on like ASASSN-14dx.
Therefore, a complete classification based on the {\it Gaia} data may be essentially difficult.

\subsection{Distances of WZ Sge-type DNe}\label{sec:dist}
We calculated the probability distributions of the distances $d$ of known WZ Sge stars
by using the {\it Gaia} parallaxes $\varpi$ and the method of \citet{bai18distance}, assumes a Galaxy model.
We obtained the median values and the 16th and 84th percentiles of the distributions.
Those of nearby object are listed in table \ref{tab:wzsge}.
ASASSN-14dx seems to be the third nearest WZ Sge-type DN.
This fact suggests that the superoutburst maximum of ASASSN-14dx can reach $V=$ 8--9 mag.
Thus ASASSN-14dx is one of the best targets for the observational research of WZ Sge-type DNe.

\begin{table}[htb]
    \caption{List of nearby WZ Sge-type DNe}
  \label{tab:wzsge}
\begin{center}
    \begin{tabular}{ccc} \hline
Object & $d$\commentc  &   $G$ \commente \\ \hline\hline
ASASSN-14dx &  80.8 [80.6, 81.1] & 15.0 \\ \hline
WZ Sge  & 45.1 [45.0, 45.2] & 15.2 \\ 
V455 And & 75.4 [75.1, 75.7] & 16.1 \\ 
BW Scl & 94.2 [93.3, 95.0] & 16.3 \\ 
V627 Peg & 99.4 [98.8, 100.1] & 15.7 \\ 
GW Lib & 112.5 [111.4, 113.5] & 16.5 \\ 
EZ Lyn & 145.3 [142.1, 148.6] & 17.8 \\ 
V355 UMa & 149.6 [147.7, 151.5] & 17.4 \\ 
FL Ps & 153.1 [150.0, 156.3] & 17.5 \\ 
V406 Vir & 168.8 [164.5, 173.3] & 17.7 \\ 
PNV J1714 \commenta & 176.9 [172.5, 181.6] & 17.2 \\ 
EG Cnc & 184.9 [175.0, 195.9] & 18.8 \\ 
QZ Lib & 189.2 [178.0, 201.9] & 18.9 \\ 
V1838 Aql & 201.6 [195.0, 208.7] & 17.9 \\ 
1RXS J0232 \commenta  & 206.9 [198.8, 215.8] & 18.7 \\ 
OV Boo \commentb & 210.5 [204.4, 216.9] & 18.2 \\ 
V624 Peg  & 217.7 [204.1, 233.2] & 18.4 \\ 
TCP J1815 \commenta  & 220.0 [208.6, 232.7] & 19.2 \\ 
SSS J1222 \commenta  & 239.8 [220.5, 262.8] & 18.9 \\ 
PNV J0309 \commenta & 246.6 [230.8, 264.6] & 18.7 \\ 
ASASSN-14cl & 261.2 [248.9, 274.9] & 18.2 \\ \hline
\hline
\multicolumn{3}{l}{\parbox{220pt}{\commentc\hspace{0.5mm} Distance estimated 
by using the parallax $\varpi$ in {\it Gaia} DR2 and
the method of \citet{bai18distance}. 
The left values are the medians of the distance probability distributions, 
and the right values in the brackets are
the 16th and 84th percentiles.
Unit of pc.
}}\\ 
\multicolumn{3}{l}{\commente\hspace{0.5mm} $G$ magnitude in {\it Gaia} DR2. Unit of mag.}\\
\multicolumn{3}{l}{\parbox{220pt}{\commenta \hspace{0.5mm}
PNV J1714 = PNV J17144255-2943481, 
1RXS J0232 = 1RXS J023238.8-371812,
TCP J1815 = TCP J18154219+3515598, 
SSS J1222 = SSS J122221.7-311523,
PNV J0309 = PNV J03093063+2638031.}}\\
\multicolumn{3}{l}{\parbox{220pt}{ \commentb\hspace{0.5mm} Population II CV
below the period minimum which showed a WZ Sge-type superoutburst.}}\\
    \end{tabular}
\end{center}
\end{table}

\section{Summary}\label{sec:summary}
Although ASASSN-14dx was classified as an SU UMa-type DN due to
the long observational gap, the object is likely to be the
best candidate of a WZ Sge-type one judging from the outburst behavior,
{\it Gaia} absolute magnitude and {\it Gaia} color.
The double-wave modulations observed in the outburst seem to be orbital modulations,
are often seen in WZ Sge stars, rather than superhumps.
The estimated distances of known WZ Sge stars indicate
that ASASSN-14dx is the third nearest WZ Sge star
and that the superoutburst maximum of ASASSN-14dx can reach $V=$ 8--9 mag.
ASASSN-14dx is one of the best targets for observational research of WZ Sge stars
because such a bright object will give us high $S/N$ data.

We examined a new classification method of WZ Sge/SU UMa-type DNe based on the {\it Gaia} DR2.
The logistic regression model provided the probability of WZ Sge-type DNe $P_{\rm WZ}$.
We proposed $P_{\rm WZ}=0.5$ as a classifier
which correctly identified 32 in 41 WZ Sge stars and 219 in 225 SU UMa ones.
A sample R code to estimate $P_{\rm WZ}$ and the galactic extinctions
is provided on our website\footnotemark[3].
The next {\it Gaia} data release and further observational studies
to increase the statistics will improve our classification method.
Our classifier will be a powerful tool to judge a priority target.

\section*{Supporting information}
The following supplementary information is available in the online article:
table E1, E2, and E3.

\section*{Acknowledgments}
This work was supported by the Grant-in-Aid for JSPS Fellows (No. 17J10039).
We are grateful to the survey projects CRTS and ASAS-SN.
This work has made use of data from the European Space Agency (ESA) mission
{\it Gaia} (https://www.cosmos.esa.int/gaia), processed by the {\it Gaia}
Data Processing and Analysis Consortium (DPAC,
https://www.cosmos.esa.int/web/gaia/dpac/consortium). Funding for the DPAC
has been provided by national institutions, in particular the institutions
participating in the {\it Gaia} Multilateral Agreement.
This research has made use of the NASA/IPAC Extragalactic Database (NED),
which is operated by the Jet Propulsion Laboratory, California Institute of Technology,
under contract with the National Aeronautics and Space Administration.
We are also thankful to the AAVSO International Database contributed by many worldwide observers.
\bibliography{pasjadd,cvs,add}
\bibliographystyle{pasjtest1}

\clearpage

\setcounter{table}{0}

\setcounter{table}{0}
\begin{table*}
\caption{Log of observations of ASASSN-14dx}
\label{tab:obs}
\begin{center}
\begin{tabular}{cccccc}
\hline
Start \commenta & End \commenta & Mag\commentb & $sigma_{Mag}$\commentc & $N$\commentd & Obs\commente \\
\hline
56854.8229 & 56854.9266 & 14.253 & 0.005 & 38 & HaC \\
56855.8201 & 56855.9271 & 14.262 & 0.005 & 39 & HaC \\
56856.8167 & 56856.9272 & 14.297 & 0.003 & 49 & HaC \\
56857.8139 & 56857.9265 & 14.282 & 0.003 & 50 & HaC \\
56858.8112 & 56858.9337 & 14.294 & 0.003 & 54 & HaC \\
56859.5509 & 56859.6768 & 14.315 & 0.002 & 363 & MLF \\
56859.8084 & 56859.9330 & 14.307 & 0.004 & 55 & HaC \\
56860.8056 & 56860.9328 & 14.304 & 0.004 & 56 & HaC \\
56861.8034 & 56861.8096 & 14.373 & 0.020 & 4 & HaC \\
56862.8009 & 56862.9326 & 14.355 & 0.003 & 58 & HaC \\
56863.7981 & 56863.9319 & 14.360 & 0.003 & 59 & HaC \\
56864.7953 & 56864.9323 & 14.379 & 0.003 & 49 & HaC \\
56865.7925 & 56865.9323 & 14.403 & 0.003 & 50 & HaC \\
56866.7897 & 56866.9305 & 14.404 & 0.004 & 51 & HaC \\
56867.7870 & 56867.9299 & 14.405 & 0.004 & 51 & HaC \\
56868.7844 & 56868.9290 & 14.415 & 0.003 & 53 & HaC \\
56869.7814 & 56869.9299 & 14.377 & 0.004 & 53 & HaC \\
56870.7787 & 56870.9299 & 14.417 & 0.003 & 54 & HaC \\
56871.7759 & 56871.9301 & 14.413 & 0.004 & 55 & HaC \\
56872.7731 & 56872.9294 & 14.424 & 0.005 & 56 & HaC \\
56873.7703 & 56873.9290 & 14.468 & 0.004 & 57 & HaC \\
56874.7675 & 56874.9293 & 14.451 & 0.004 & 58 & HaC \\
56875.7648 & 56875.9271 & 14.501 & 0.003 & 58 & HaC \\
56876.7620 & 56876.9269 & 14.483 & 0.003 & 59 & HaC \\
56879.7537 & 56879.9099 & 14.514 & 0.005 & 56 & HaC \\
56880.7509 & 56880.9092 & 14.505 & 0.005 & 57 & HaC \\
56881.7473 & 56881.9103 & 14.516 & 0.003 & 78 & HaC \\
56882.7454 & 56882.9090 & 14.520 & 0.003 & 78 & HaC \\
56885.7655 & 56885.9145 & 14.479 & 0.005 & 49 & HaC \\
56886.7627 & 56886.9115 & 14.509 & 0.004 & 56 & HaC \\
56887.7599 & 56887.9091 & 14.511 & 0.004 & 49 & HaC \\
56888.7572 & 56888.9092 & 14.512 & 0.004 & 50 & HaC \\
56889.7544 & 56889.8950 & 14.516 & 0.004 & 45 & HaC \\
56890.7516 & 56890.9099 & 14.535 & 0.003 & 52 & HaC \\
56891.7488 & 56891.9070 & 14.552 & 0.003 & 52 & HaC \\
56892.7461 & 56892.8787 & 14.547 & 0.004 & 43 & HaC \\
56893.7433 & 56893.8719 & 14.548 & 0.004 & 42 & HaC \\
56894.7412 & 56894.8815 & 14.555 & 0.005 & 46 & HaC \\
56895.7384 & 56895.9052 & 14.565 & 0.004 & 55 & HaC \\
56896.7357 & 56896.9058 & 14.561 & 0.004 & 56 & HaC \\
56897.7329 & 56897.9055 & 14.584 & 0.004 & 57 & HaC \\
56898.7121 & 56898.9034 & 14.554 & 0.004 & 64 & HaC \\
56900.7245 & 56900.9039 & 14.548 & 0.004 & 59 & HaC \\
56901.7217 & 56901.9030 & 14.587 & 0.003 & 60 & HaC \\
\hline
  \multicolumn{6}{l}{\commenta BJD$-$2400000.} \\
  \multicolumn{6}{l}{\commentb Mean magnitude. All observations are no filter (clear).} \\
  \multicolumn{6}{l}{\commentb Standard deviation of the observed magnitude.} \\
  \multicolumn{6}{l}{\commentd Number of observations.} \\
  \multicolumn{6}{l}{\commente Observer's code: HaC(F. J. Hambsch), MLF (B. Monard),}\\
  \multicolumn{6}{l}{ KU1 (Kyoto U. team), Kis (S. Kiyota)}\\
\end{tabular}
\end{center}
\end{table*}

\setcounter{table}{0}
\begin{table*}
\caption{Log of observations of ASASSN-14dx (continued)}
\label{tab:obs}
\begin{center}
\begin{tabular}{cccccc}
\hline
Start \commenta & End \commenta & Mag\commentb & $sigma_{Mag}$\commentc & $N$\commentd & Obs\commente \\
\hline
56902.7190 & 56902.9003 & 14.592 & 0.004 & 60 & HaC \\
56903.7495 & 56903.8999 & 14.575 & 0.005 & 50 & HaC \\
56904.7230 & 56904.7815 & 14.575 & 0.007 & 20 & HaC \\
56905.7456 & 56905.8983 & 14.576 & 0.004 & 51 & HaC \\
56906.7442 & 56906.8995 & 14.606 & 0.004 & 52 & HaC \\
56907.8263 & 56907.9158 & 14.626 & 0.006 & 37 & HaC \\
56908.7389 & 56908.8979 & 14.630 & 0.005 & 54 & HaC \\
56909.7361 & 56909.8972 & 14.648 & 0.005 & 55 & HaC \\
56911.7306 & 56911.8933 & 14.627 & 0.005 & 54 & HaC \\
56913.7249 & 56913.8583 & 14.610 & 0.009 & 38 & HaC \\
56914.7228 & 56914.8909 & 14.623 & 0.004 & 60 & HaC \\
56915.7200 & 56915.8900 & 14.595 & 0.003 & 61 & HaC \\
56916.7173 & 56916.8892 & 14.620 & 0.004 & 62 & HaC \\
56919.7089 & 56919.8871 & 14.646 & 0.004 & 65 & HaC \\
56920.7061 & 56920.8864 & 14.638 & 0.004 & 66 & HaC \\
56921.7033 & 56921.8856 & 14.631 & 0.004 & 67 & HaC \\
56922.7006 & 56922.8849 & 14.639 & 0.003 & 68 & HaC \\
56923.6978 & 56923.8844 & 14.626 & 0.006 & 68 & HaC \\
56924.6951 & 56924.8836 & 14.647 & 0.004 & 70 & HaC \\
56925.6923 & 56925.8821 & 14.646 & 0.005 & 69 & HaC \\
56926.6895 & 56926.8821 & 14.632 & 0.005 & 72 & HaC \\
56927.6867 & 56927.8818 & 14.667 & 0.004 & 74 & HaC \\
56928.6839 & 56928.8772 & 14.655 & 0.004 & 72 & HaC \\
56929.6811 & 56929.8764 & 14.665 & 0.004 & 74 & HaC \\
56930.6783 & 56930.8735 & 14.678 & 0.006 & 74 & HaC \\
56931.6756 & 56931.8789 & 14.696 & 0.004 & 78 & HaC \\
56934.8403 & 56934.8701 & 14.648 & 0.005 & 16 & HaC \\
56935.6652 & 56935.8720 & 14.702 & 0.003 & 79 & HaC \\
56936.6624 & 56936.8713 & 14.720 & 0.004 & 80 & HaC \\
56937.6596 & 56937.8701 & 14.717 & 0.004 & 81 & HaC \\
56938.6569 & 56938.8699 & 14.751 & 0.006 & 89 & HaC \\
56939.6540 & 56939.8703 & 14.706 & 0.005 & 87 & HaC \\
56940.6680 & 56940.8624 & 14.699 & 0.008 & 79 & HaC \\
56941.6652 & 56941.8679 & 14.723 & 0.005 & 71 & HaC \\
56942.7241 & 56942.8669 & 14.724 & 0.006 & 51 & HaC \\
56943.6597 & 56943.8652 & 14.730 & 0.005 & 72 & HaC \\
56944.6569 & 56944.8654 & 14.751 & 0.004 & 73 & HaC \\
56945.6569 & 56945.8659 & 14.748 & 0.004 & 72 & HaC \\
56946.6513 & 56946.8628 & 14.760 & 0.004 & 73 & HaC \\
56947.6492 & 56947.8599 & 14.768 & 0.006 & 74 & HaC \\
56948.6464 & 56948.8607 & 14.780 & 0.004 & 75 & HaC \\
56949.6436 & 56949.8581 & 14.785 & 0.005 & 75 & HaC \\
56950.6408 & 56950.8524 & 14.750 & 0.003 & 74 & HaC \\
56951.6380 & 56951.8478 & 14.798 & 0.005 & 69 & HaC \\
56953.6321 & 56953.8505 & 14.865 & 0.004 & 76 & HaC \\
56954.6856 & 56954.8491 & 14.882 & 0.006 & 57 & HaC \\
56955.6828 & 56955.8494 & 14.854 & 0.005 & 58 & HaC \\
56957.3066 & 56957.3752 & 14.864 & 0.002 & 198 & MLF \\
56959.6718 & 56959.8469 & 14.889 & 0.006 & 65 & HaC \\
56960.6689 & 56960.8466 & 14.893 & 0.005 & 66 & HaC \\
\hline
\end{tabular}
\end{center}
\end{table*}

\setcounter{table}{0}
\begin{table*}
\caption{Log of observations of ASASSN-14dx (continued)}
\label{tab:obs}
\begin{center}
\begin{tabular}{cccccc}
\hline
Start \commenta & End \commenta & Mag\commentb & $sigma_{Mag}$\commentc & $N$\commentd & Obs\commente \\
\hline
56961.6662 & 56961.8438 & 14.888 & 0.005 & 66 & HaC \\
56962.6634 & 56962.8444 & 14.887 & 0.005 & 72 & HaC \\
56963.6607 & 56963.8436 & 14.899 & 0.006 & 73 & HaC \\
56964.6579 & 56964.8429 & 14.870 & 0.004 & 74 & HaC \\
56965.7484 & 56965.8428 & 14.864 & 0.006 & 44 & HaC \\
56966.6530 & 56966.8416 & 14.886 & 0.005 & 76 & HaC \\
56967.6647 & 56967.8408 & 14.915 & 0.008 & 76 & HaC \\
56968.6625 & 56968.8422 & 14.876 & 0.007 & 77 & HaC \\
56969.6597 & 56969.8317 & 14.846 & 0.008 & 79 & HaC \\
56970.5876 & 56970.8292 & 14.808 & 0.004 & 96 & HaC \\
56971.6603 & 56971.8280 & 14.846 & 0.005 & 62 & HaC \\
56972.6540 & 56972.8281 & 14.830 & 0.004 & 65 & HaC \\
56973.6512 & 56973.8279 & 14.823 & 0.004 & 66 & HaC \\
56974.6484 & 56974.8278 & 14.848 & 0.004 & 67 & HaC \\
56975.6457 & 56975.8278 & 14.849 & 0.004 & 68 & HaC \\
56976.6429 & 56976.8278 & 14.921 & 0.005 & 69 & HaC \\
56977.6401 & 56977.8251 & 14.934 & 0.004 & 69 & HaC \\
56979.6346 & 56979.8246 & 14.944 & 0.004 & 71 & HaC \\
56980.6619 & 56980.8250 & 14.929 & 0.003 & 60 & HaC \\
56981.6591 & 56981.8250 & 14.941 & 0.004 & 62 & HaC \\
56982.6563 & 56982.8249 & 14.924 & 0.004 & 63 & HaC \\
56983.6535 & 56983.8247 & 14.951 & 0.004 & 64 & HaC \\
56984.6507 & 56984.7856 & 14.941 & 0.003 & 54 & HaC \\
56985.6528 & 56985.7813 & 14.925 & 0.004 & 46 & HaC \\
56986.6907 & 56986.7788 & 14.920 & 0.004 & 31 & HaC \\
56987.6492 & 56987.7766 & 14.940 & 0.004 & 45 & HaC \\
56988.5774 & 56988.6999 & 14.899 & 0.002 & 316 & KU1 \\
56988.6464 & 56988.7732 & 14.930 & 0.003 & 45 & HaC \\
56989.6436 & 56989.7702 & 14.913 & 0.004 & 45 & HaC \\
56990.6408 & 56990.7677 & 14.925 & 0.004 & 45 & HaC \\
56991.6381 & 56991.7655 & 14.930 & 0.005 & 45 & HaC \\
56992.6353 & 56992.7627 & 14.932 & 0.003 & 45 & HaC \\
56993.6325 & 56993.7600 & 14.951 & 0.006 & 45 & HaC \\
56994.6297 & 56994.7580 & 14.928 & 0.007 & 56 & HaC \\
56995.6442 & 56995.7552 & 14.946 & 0.007 & 48 & HaC \\
56997.6423 & 56997.7504 & 14.952 & 0.011 & 41 & HaC \\
56998.6403 & 56998.7476 & 14.933 & 0.008 & 41 & HaC \\
56999.6374 & 56999.7449 & 14.935 & 0.005 & 41 & HaC \\
57000.6346 & 57000.7399 & 14.981 & 0.008 & 40 & HaC \\
57002.6844 & 57002.7344 & 14.945 & 0.007 & 21 & HaC \\
57003.6817 & 57003.7316 & 14.937 & 0.007 & 21 & HaC \\
57004.6789 & 57004.7290 & 14.961 & 0.008 & 21 & HaC \\
57005.6762 & 57005.7262 & 14.914 & 0.014 & 21 & HaC \\
57007.6331 & 57007.7219 & 14.935 & 0.004 & 53 & HaC \\
57008.6678 & 57008.7187 & 14.905 & 0.008 & 24 & HaC \\
57013.8881 & 57014.0847 & 14.950 & 0.003 & 520 & Kis \\
57317.0156 & 57317.0932 & 15.347 & 0.007 & 196 & KU1 \\
57317.9675 & 57318.1806 & 15.332 & 0.003 & 561 & KU1 \\
\hline
\end{tabular}
\end{center}
\end{table*}

\setcounter{table}{1}
\begin{table*}
\caption{List of WZ Sge-type DNe and candidates identified with $Gaia$ DR2 objects}
\begin{center}
\scalebox{0.9}{
\begin{tabular}{cccccccc}
\hline
Name & Plx\commenta & $G$\commentb & $G_{\rm col}$\commentc & $A_{G}$\commentd & $A_{\rm BP} - A_{\rm RP}$\commentd & $P_{\rm orb}$\commente & $P_{\rm WZ}$\commenth \\
\hline
SDSS J160501.35+203056.9 & 3.11(55) & 19.822(6) & 0.066(87) & 0.179(14) & 0.100(18) & 0.05666 & 0.97(2) \\ 
OT J012059.59 +325545.0 & 2.80(50) & 19.852(6) & 0.018(124) & 0.118(13) & 0.066(17) & 0.05716 & 0.97(3) \\ 
EG Cnc & 5.44(31) & 18.812(2) & 0.175(35) & 0.085(4) & 0.048(6) & 0.05997 & 0.96(1) \\ 
QZ Lib & 5.34(33) & 18.880(10) & 0.262(62) & 0.222(13) & 0.124(17) & 0.06413 & 0.95(3) \\ 
DY CMi & 3.29(42) & 19.505(3) & 0.099(52) & 0.043(6) & 0.024(8) & 0.05937\commentf & 0.94(3) \\ 
CRTS J224739.7-362254 & 3.22(52) & 19.649(5) & 0.131(67) & 0.033(2) & 0.019(2) & 0.05572\commentf & 0.94(5) \\ 
TCP J18154219+3515598 & 4.57(25) & 19.193(4) & 0.240(69) & 0.065(4) & 0.037(5) & 0.06022\commentf & 0.94(3) \\ 
PQ And & 3.98(45) & 19.063(5) & 0.126(69) & 0.085(11) & 0.047(14) & 0.05600 & 0.93(4) \\ 
1RXS J023238.8-371812 & 4.83(20) & 18.675(6) & 0.150(46) & 0.068(2) & 0.038(2) & 0.06500 & 0.93(2) \\ 
EZ Lyn & 6.87(15) & 17.810(2) & 0.128(19) & 0.078(2) & 0.044(3) & 0.05900 & 0.92(1) \\ 
UZ Boo & 3.62(61) & 19.992(6) & 0.383(77) & 0.068(4) & 0.038(6) & 0.06051\commentf & 0.91(7) \\ 
WZ Sge & 22.16(4) & 15.210(3) & 0.153(19) & 0.048(0) & 0.027(0) & 0.05669 & 0.90(1) \\ 
V624 Peg & 4.64(31) & 18.424(5) & 0.159(31) & 0.177(12) & 0.098(17) & 0.05865 & 0.88(4) \\ 
PNV J03093063+2638031 & 4.09(28) & 18.652(15) & 0.214(79) & 0.339(21) & 0.189(28) & 0.05615 & 0.86(7) \\ 
V355 UMa & 6.66(9) & 17.383(2) & 0.082(15) & 0.018(0) & 0.010(0) & 0.05729 & 0.85(1) \\ 
SSS J122221.7-311523 & 4.26(36) & 18.851(12) & 0.266(75) & 0.119(9) & 0.066(12) & 0.07588 & 0.84(8) \\ 
SDSS J161027.61+090738.4 & 2.63(51) & 19.838(16) & 0.250(125) & 0.096(7) & 0.053(10) & 0.05687 & 0.83(16) \\ 
ASASSN-14cv & 3.43(21) & 19.004(16) & 0.179(102) & 0.071(3) & 0.040(4) & 0.05992 & 0.82(11) \\ 
V406 Vir & 5.91(16) & 17.718(7) & 0.152(35) & 0.061(1) & 0.034(2) & 0.05592 & 0.82(4) \\ 
ASASSN-14jv & 3.52(18) & 18.697(18) & 0.141(95) & 0.164(9) & 0.091(12) & 0.05442 & 0.81(10) \\ 
V1838 Aql & 4.95(17) & 17.866(9) & 0.188(47) & 0.362(20) & 0.202(26) & 0.05706 & 0.79(6) \\ 
ASAS J102522-1542.4 & 3.91(52) & 19.357(15) & 0.438(86) & 0.160(16) & 0.090(22) & 0.06136 & 0.76(15) \\ 
BW Scl & 10.60(10) & 16.262(4) & 0.164(26) & 0.019(0) & 0.011(0) & 0.05432 & 0.72(4) \\ 
GW Lib & 8.87(8) & 16.490(7) & 0.166(28) & 0.175(2) & 0.098(3) & 0.05332 & 0.72(4) \\ 
FL Psc & 6.51(14) & 17.527(6) & 0.279(27) & 0.165(3) & 0.092(4) & 0.05610 & 0.70(4) \\ 
OV Boo & 4.74(14) & 18.234(6) & 0.259(49) & 0.038(1) & 0.021(1) & 0.04626 & 0.69(8) \\ 
PNV J17144255-2943481 & 5.64(15) & 17.163(7) & 0.151(52) & 0.249(10) & 0.139(13) & 0.05956 & 0.64(9) \\ 
ASASSN-17el & 2.91(17) & 18.680(4) & 0.131(43) & 0.071(3) & 0.040(4) & 0.05516 & 0.63(9) \\ 
HV Vir & 2.90(41) & 19.138(9) & 0.276(61) & 0.083(3) & 0.046(4) & 0.05707 & 0.61(17) \\ 
PNV J23052314-0225455 & 2.78(42) & 19.047(25) & 0.249(122) & 0.152(7) & 0.085(9) & 0.05456 & 0.60(25) \\ 
EQ Lyn & 2.70(39) & 18.968(6) & 0.276(37) & 0.129(11) & 0.072(16) & 0.05278 & 0.48(16) \\ 
V455 And & 13.24(6) & 16.064(4) & 0.419(21) & 0.064(1) & 0.036(1) & 0.05631 & 0.47(4) \\ 
ASASSN-14cl & 3.83(19) & 18.187(13) & 0.400(53) & 0.091(5) & 0.051(6) & 0.05838 & 0.26(8) \\ 
SDSS J145758.21+514807.9 & 1.72(31) & 19.821(13) & 0.435(111) & 0.067(2) & 0.038(1) & 0.05409 & 0.18(16) \\ 
MASTER OT J085854.16-274030.7 & 2.03(35) & 19.092(6) & 0.381(55) & 0.219(36) & 0.123(49) & 0.05475\commentf & 0.17(12) \\ 
BC UMa & 3.12(18) & 18.370(4) & 0.427(29) & 0.056(1) & 0.031(1) & 0.06261 & 0.14(4) \\ 
V627 Peg & 10.03(7) & 15.667(4) & 0.380(37) & 0.050(1) & 0.028(1) & 0.05452 & 0.14(3) \\ 
ASASSN-16eg & 2.25(32) & 19.506(8) & 0.694(67) & 0.049(3) & 0.027(4) & 0.07548 & 0.06(4) \\ 
RZ Leo & 3.56(25) & 18.241(12) & 0.791(50) & 0.058(2) & 0.032(2) & 0.07604 & 0.02(1) \\ 
\hline
    \multicolumn{8}{c}{WZ Sge-type DN candidates.} \\
\hline
SSS J200331.3-284938 & 3.10(47) & 19.564(18) & 0.380(107) & 0.195(21) & 0.109(29) & 0.05871 & 0.74(19) \\ 
PNV J19321040-2052505 & 3.00(50) & 19.141(8) & 0.298(46) & 0.130(20) & 0.073(28) & --- & 0.63(18) \\ 
ASASSN-17kg & 3.31(51) & 19.147(8) & 0.366(41) & 0.109(7) & 0.061(9) & 0.05660\commentf & 0.62(17) \\ 
ASASSN-14dx & 12.34(4) & 14.959(25) & 0.232(127) & 0.034(1) & 0.019(1) & 0.05751 & 0.21(14) \\ 
ASASSN-14dx in quiescence\commentg & 12.34(4) & 16.254(42) & -0.112(74)  & 0.037(0) & 0.020(0) & 0.05751 & 0.97(1) \\ 
\hline
\multicolumn{8}{l}{\commenta Parallax $\varpi$ obtained by {\it Gaia} DR2 in units of mas. } \\
  \multicolumn{8}{l}{\commentb {\it Gaia} $G$ magnitude} \\
  \multicolumn{8}{l}{\commentc {\it Gaia} color $G_{\rm col} = G_{\rm BP}-G_{\rm RP}$ in units of magnitude} \\
  \multicolumn{8}{l}{\commentd Galactic extinction in $G$, $G_{\rm BP}$ and $G_{\rm RP}$ in units of magnitude. The estimation method is described in section 4.2.} \\
  \multicolumn{8}{l}{\commente  Orbital period in units of day} \\
  \multicolumn{8}{l}{\parbox{400pt}{\commenth  WZ Sge-type DN probability $P_{\rm WZ}$.
$P_{\rm WZ}=0.5$ is a borderline between WZ Sge/SU UMa stars.
Objects with $P_{\rm WZ} > 0.5$ are good candidates of WZ Sge stars.}} \\
  \multicolumn{8}{l}{\commentg $G$ and $G_{\rm col}$ in quiescence are calculated by using the SDSS DR9.} \\
  \multicolumn{8}{l}{\commentf Orbital periods estimated from superhump ones by using equation (6) in Kato et al. (2012a). } \\
\end{tabular}
}
\end{center}
\end{table*}

\setcounter{table}{2}
\begin{table*}
\caption{List of SU UMa-type DNe identified with $Gaia$ DR2 objects}
\begin{center}
\scalebox{0.9}{
\begin{tabular}{cccccccc}
\hline
Name & Plx\commenta & $G$\commentb & $G_{\rm col}$\commentc & $A_{G}$\commentd & $A_{\rm BP} - A_{\rm RP}$\commentd & $P_{\rm orb}$\commente & $P_{\rm WZ}$\commenth \\
\hline 
MASTER OT J220559.40-341434.9 & 6.23(27) & 18.344(5) & 0.157(30) & 0.045(2) & 0.025(2) & 0.06128 & 0.95(1) \\ 
MASTER OT J143857.37+703523.1 & 3.82(26) & 19.371(4) & 0.146(49) & 0.035(2) & 0.019(2) & --- & 0.94(2) \\ 
ASASSN-15lt & 2.50(48) & 19.696(23) & 0.193(186) & 0.192(20) & 0.107(26) & --- & 0.84(20) \\ 
SDSS J150137.22+550123.4 & 2.93(28) & 19.257(5) & 0.144(56) & 0.030(1) & 0.017(1) & 0.05684 & 0.82(8) \\ 
XZ Eri & 3.00(30) & 19.247(9) & 0.275(96) & 0.075(4) & 0.042(5) & 0.06116 & 0.69(16) \\ 
ASASSN-17eq & 2.81(38) & 19.956(25) & 0.519(149) & 0.059(4) & 0.033(5) & --- & 0.55(28) \\ 
ASASSN-15ev & 2.26(38) & 19.574(16) & 0.423(83) & 0.410(38) & 0.231(52) & --- & 0.47(24) \\ 
ASASSN-15fo & 2.44(41) & 19.474(11) & 0.470(101) & 0.157(24) & 0.088(33) & --- & 0.31(21) \\ 
MM Hya & 2.77(23) & 18.701(17) & 0.324(88) & 0.105(6) & 0.059(8) & 0.05759 & 0.30(15) \\ 
GP CVn & 2.75(20) & 18.791(12) & 0.344(74) & 0.039(1) & 0.022(1) & 0.06295 & 0.27(12) \\ 
DT Pyx & 2.88(29) & 19.012(8) & 0.482(49) & 0.201(21) & 0.113(30) & 0.06165 & 0.27(11) \\ 
NZ Boo & 5.36(9) & 17.342(10) & 0.363(54) & 0.030(0) & 0.017(0) & 0.05891 & 0.24(7) \\ 
SDSS J090350.73+330036.1 & 2.52(38) & 19.064(17) & 0.422(87) & 0.055(3) & 0.031(4) & 0.05907 & 0.21(14) \\ 
TV Crv & 3.07(30) & 18.679(7) & 0.467(43) & 0.163(7) & 0.092(10) & 0.06288 & 0.21(8) \\ 
ASASSN-14hl & 1.21(23) & 19.929(14) & 0.219(118) & 0.035(1) & 0.020(0) & --- & 0.20(18) \\ 
WX Cet & 3.79(16) & 18.137(7) & 0.427(40) & 0.044(1) & 0.025(1) & 0.05826 & 0.19(5) \\ 
VY Aqr & 7.24(14) & 16.865(27) & 0.519(116) & 0.181(5) & 0.102(6) & 0.06309 & 0.17(11) \\ 
ASASSN-17kc & 1.95(35) & 19.196(10) & 0.367(72) & 0.127(5) & 0.071(6) & --- & 0.17(12) \\ 
MASTER OT J003831.10-640313.7 & 3.00(16) & 18.312(11) & 0.369(61) & 0.040(1) & 0.022(1) & --- & 0.15(6) \\ 
ASASSN-17qn & 2.49(44) & 19.819(14) & 0.731(120) & 0.192(23) & 0.108(33) & --- & 0.15(14) \\ 
CRTS J215815.3+094709 & 3.70(16) & 17.488(19) & 0.291(94) & 0.113(4) & 0.063(5) & --- & 0.15(8) \\ 
OY Car & 11.01(3) & 15.620(34) & 0.427(145) & 0.077(2) & 0.043(1) & 0.06312 & 0.15(12) \\ 
UV Per & 3.99(13) & 17.926(8) & 0.488(48) & 0.169(9) & 0.095(12) & 0.06489 & 0.14(4) \\ 
FZ Cet & 2.89(17) & 18.244(26) & 0.347(128) & 0.030(1) & 0.017(1) & --- & 0.14(10) \\ 
ASASSN-17fo & 2.51(43) & 19.158(16) & 0.528(97) & 0.051(2) & 0.029(2) & 0.06155 & 0.13(11) \\ 
ASASSN-15sp & 1.84(35) & 19.789(20) & 0.660(131) & 0.493(72) & 0.279(102) & --- & 0.12(14) \\ 
QZ Vir & 7.81(7) & 16.055(12) & 0.363(50) & 0.036(0) & 0.020(0) & 0.05882 & 0.12(4) \\ 
V359 Cen & 2.69(23) & 18.753(12) & 0.539(67) & 0.261(18) & 0.147(26) & 0.07790 & 0.12(6) \\ 
CC Scl & 4.90(15) & 16.882(10) & 0.323(48) & 0.032(1) & 0.018(1) & 0.05857 & 0.11(3) \\ 
ASASSN-15mb & 2.45(21) & 19.066(15) & 0.525(85) & 0.043(1) & 0.024(1) & --- & 0.11(6) \\ 
SW UMa & 6.15(8) & 16.576(10) & 0.391(40) & 0.058(1) & 0.033(1) & 0.05682 & 0.10(3) \\ 
CRTS J034515.4-015216 & 2.61(37) & 18.930(16) & 0.606(114) & 0.313(17) & 0.176(23) & 0.07018 & 0.10(9) \\ 
MASTER OT J004527.52+503213.8 & 3.29(32) & 18.974(7) & 0.755(58) & 0.227(26) & 0.128(37) & 0.07800 & 0.09(5) \\ 
MASTER OT J072948.66+593824.4 & 2.70(51) & 19.000(23) & 0.605(86) & 0.104(12) & 0.058(17) & --- & 0.09(8) \\ 
ASASSN-18gn & 2.26(43) & 19.764(12) & 0.755(95) & 0.228(43) & 0.129(61) & --- & 0.09(9) \\ 
IY UMa & 5.60(10) & 17.438(23) & 0.596(84) & 0.019(0) & 0.010(0) & 0.07391 & 0.09(5) \\ 
2QZ J015940.6-281040 & 2.58(37) & 19.157(9) & 0.609(105) & 0.029(1) & 0.016(1) & --- & 0.09(7) \\ 
BC Dor & 2.63(16) & 18.813(9) & 0.566(64) & 0.157(5) & 0.088(7) & 0.06613 & 0.08(4) \\ 
SDSS J140037.99+572341.3 & 1.81(26) & 18.878(18) & 0.327(96) & 0.034(1) & 0.019(1) & 0.06300 & 0.08(7) \\ 
NSV 4618 & 5.47(12) & 16.851(18) & 0.442(69) & 0.070(2) & 0.039(2) & 0.06577 & 0.08(3) \\ 
ASASSN-17gf & 2.07(17) & 18.517(12) & 0.409(74) & 0.396(41) & 0.222(56) & --- & 0.08(5) \\ 
V521 Peg & 5.23(9) & 17.044(7) & 0.495(38) & 0.083(1) & 0.047(2) & 0.05990 & 0.07(2) \\ 
CRTS J044027.1+023301 & 2.31(26) & 18.673(19) & 0.550(92) & 0.414(25) & 0.234(34) & --- & 0.07(5) \\ 
GO Com & 2.64(24) & 18.324(13) & 0.438(67) & 0.022(0) & 0.012(0) & 0.06580 & 0.06(3) \\ 
CRTS J164950.4+035835 & 2.33(26) & 18.649(10) & 0.504(44) & 0.200(11) & 0.112(15) & --- & 0.06(3) \\ 
\hline
\multicolumn{8}{l}{\commenta Parallax $\varpi$ obtained by {\it Gaia} DR2 in units of mas. } \\
  \multicolumn{8}{l}{\commentb {\it Gaia} $G$ magnitude} \\
  \multicolumn{8}{l}{\commentc {\it Gaia} color $G_{\rm col} = G_{\rm BP}-G_{\rm RP}$ in units of magnitude} \\
  \multicolumn{8}{l}{\commentd Galactic extinction in $G$, $G_{\rm BP}$ and $G_{\rm RP}$ in units of magnitude. The estimation method is described in section 4.2.} \\
  \multicolumn{8}{l}{\commente  Orbital period in units of day} \\
  \multicolumn{8}{l}{\parbox{400pt}{\commenth  WZ Sge-type DN probability $P_{\rm WZ}$.
$P_{\rm WZ}=0.5$ is a borderline between WZ Sge/SU UMa stars.
Objects with $P_{\rm WZ} > 0.5$ are good candidates of WZ Sge stars.}} \\
\end{tabular}
}
\end{center}
\end{table*}

\setcounter{table}{2}
\begin{table*}
\caption{List of SU UMa-type DNe identified with $Gaia$ DR2 objects (continued)}
\begin{center}
\scalebox{0.9}{
\begin{tabular}{cccccccc}
\hline
Name & Plx\commenta & $G$\commentb & $G_{\rm col}$\commentc & $A_{G}$\commentd & $A_{\rm BP} - A_{\rm RP}$\commentd & $P_{\rm orb}$\commente & $P_{\rm WZ}$\commenth \\
\hline 
ASASSN-14kb & 2.91(19) & 18.667(32) & 0.638(187) & 0.155(7) & 0.087(8) & 0.06811 & 0.06(7) \\ 
V436 Cen & 6.36(6) & 16.323(30) & 0.446(136) & 0.114(3) & 0.064(3) & 0.06250 & 0.06(5) \\ 
PNV J06501960+3002449 & 2.64(24) & 18.443(32) & 0.514(174) & 0.137(14) & 0.077(19) & --- & 0.06(7) \\ 
SDSS J182142.83+212153.5 & 2.03(29) & 19.272(10) & 0.628(58) & 0.274(30) & 0.155(43) & --- & 0.06(4) \\ 
ASASSN-14je & 1.70(32) & 19.563(26) & 0.598(136) & 0.222(16) & 0.125(22) & 0.06700 & 0.05(6) \\ 
V844 Her & 3.41(10) & 17.489(16) & 0.396(62) & 0.020(0) & 0.011(0) & 0.05464 & 0.05(2) \\ 
OV Dra & 2.14(15) & 18.592(25) & 0.480(88) & 0.043(1) & 0.024(1) & 0.05874 & 0.04(2) \\ 
SDSS J033449.86-071047.8 & 2.33(30) & 18.746(13) & 0.605(65) & 0.112(4) & 0.063(5) & 0.07900 & 0.03(2) \\ 
BZ UMa & 6.56(6) & 16.206(17) & 0.511(63) & 0.077(1) & 0.043(1) & 0.06799 & 0.03(1) \\ 
VW Hyi & 18.53(2) & 13.837(19) & 0.484(85) & 0.077(1) & 0.044(1) & 0.07427 & 0.03(2) \\ 
MASTER OT J120251.56-454116.7 & 2.22(24) & 18.990(19) & 0.670(95) & 0.177(16) & 0.100(22) & --- & 0.03(2) \\ 
SY Cap & 2.60(18) & 18.104(10) & 0.518(44) & 0.152(6) & 0.086(9) & --- & 0.03(1) \\ 
OGLE-BLG-DN-0254 & 3.76(16) & 18.092(16) & 0.761(61) & 0.213(14) & 0.120(20) & 0.07193 & 0.03(1) \\ 
V630 Cyg & 2.22(32) & 19.193(15) & 0.808(93) & 0.404(68) & 0.229(99) & --- & 0.03(3) \\ 
AY For & 2.09(24) & 18.304(18) & 0.418(82) & 0.039(1) & 0.022(1) & 0.07460 & 0.03(2) \\ 
SDSS J131432.10+444138.7 & 1.48(25) & 19.089(15) & 0.431(92) & 0.045(1) & 0.025(1) & --- & 0.03(2) \\ 
ASASSN-17hm & 2.30(43) & 19.961(19) & 1.058(130) & 0.406(105) & 0.230(156) & --- & 0.03(4) \\ 
V1024 Per & 3.98(7) & 17.062(21) & 0.554(94) & 0.329(10) & 0.186(12) & 0.07060 & 0.03(2) \\ 
V493 Ser & 3.24(21) & 18.327(9) & 0.783(57) & 0.309(12) & 0.174(16) & 0.07408 & 0.03(1) \\ 
EG Aqr & 2.96(38) & 18.971(24) & 0.855(163) & 0.074(3) & 0.042(4) & 0.07596 & 0.02(3) \\ 
ASASSN-15au & 2.36(18) & 17.877(12) & 0.395(75) & 0.046(3) & 0.026(4) & --- & 0.02(1) \\ 
MASTER OT J064725.70+491543.9 & 3.73(15) & 17.372(20) & 0.561(95) & 0.139(6) & 0.078(8) & 0.06554 & 0.02(2) \\ 
SDSS J032015.29+441059.3 & 2.11(24) & 18.870(30) & 0.676(185) & 0.273(33) & 0.154(46) & 0.06870 & 0.02(3) \\ 
HT Cas & 7.07(6) & 16.351(11) & 0.668(48) & 0.115(2) & 0.065(2) & 0.07365 & 0.02(1) \\ 
CRTS J170609.7+143452 & 2.26(18) & 18.123(22) & 0.526(92) & 0.309(12) & 0.174(16) & 0.05823 & 0.02(2) \\ 
ASASSN-14ag & 5.63(9) & 16.180(23) & 0.465(109) & 0.043(1) & 0.024(1) & 0.06031 & 0.02(2) \\ 
BB Ari & 2.83(19) & 18.450(12) & 0.744(65) & 0.231(10) & 0.130(14) & 0.07020 & 0.02(1) \\ 
CRTS J050659.2-165932 & 1.78(23) & 18.663(14) & 0.499(81) & 0.155(6) & 0.087(8) & --- & 0.02(2) \\ 
AQ CMi & 1.55(27) & 18.675(26) & 0.434(135) & 0.159(27) & 0.089(37) & 0.06490 & 0.02(2) \\ 
CRTS J094327.3-272039 & 2.07(27) & 18.611(28) & 0.618(135) & 0.232(21) & 0.131(29) & --- & 0.02(2) \\ 
MR UMa & 2.93(17) & 17.893(7) & 0.580(44) & 0.052(1) & 0.029(1) & 0.06337 & 0.02(1) \\ 
CU Vel & 6.29(6) & 16.712(13) & 0.772(57) & 0.254(5) & 0.143(6) & 0.07850 & 0.02(1) \\ 
V1028 Cyg & 1.73(22) & 18.890(21) & 0.611(152) & 0.307(31) & 0.173(43) & --- & 0.02(2) \\ 
V2051 Oph & 8.90(7) & 15.367(9) & 0.590(45) & 0.186(3) & 0.105(3) & 0.06243 & 0.02(1) \\ 
ASASSN-13cz & 1.75(22) & 18.775(9) & 0.514(41) & 0.041(1) & 0.023(1) & 0.07700 & 0.02(1) \\ 
SDSS J110014.72+131552.1 & 2.17(29) & 18.681(42) & 0.623(196) & 0.040(2) & 0.022(1) & 0.06560 & 0.02(2) \\ 
ASASSN-13cx & 2.26(23) & 18.054(22) & 0.526(100) & 0.210(16) & 0.118(22) & 0.07965 & 0.02(1) \\ 
EF Peg & 3.92(45) & 18.080(11) & 0.853(46) & 0.108(12) & 0.061(17) & 0.08370 & 0.02(1) \\ 
CRTS J163120.9+103134 & 2.03(12) & 18.522(14) & 0.594(109) & 0.192(5) & 0.108(5) & 0.06265 & 0.02(1) \\ 
AY Lyr & 2.22(13) & 17.932(34) & 0.464(153) & 0.121(7) & 0.068(8) & 0.07330 & 0.02(2) \\ 
1RXS J185310.0+594509 & 1.53(15) & 18.489(12) & 0.396(70) & 0.102(4) & 0.057(5) & 0.05826 & 0.01(1) \\ 
ASASSN-13bj & 1.80(21) & 18.523(26) & 0.503(121) & 0.067(2) & 0.038(2) & --- & 0.01(1) \\ 
V776 And & 2.45(15) & 18.119(11) & 0.628(54) & 0.249(11) & 0.141(15) & 0.06400 & 0.01(1) \\ 
TY Psc & 3.83(10) & 16.866(26) & 0.516(113) & 0.115(3) & 0.065(4) & 0.06833 & 0.01(1) \\ 
SDSS J152419.33+220920.0 & 2.15(29) & 19.143(30) & 0.810(131) & 0.119(4) & 0.067(4) & 0.06532 & 0.01(1) \\ 
Mis V1446 & 2.01(26) & 18.906(25) & 0.714(149) & 0.172(28) & 0.097(39) & --- & 0.01(2) \\ 
MASTER OT J162323.48+782603.3 & 2.51(16) & 18.448(14) & 0.695(66) & 0.079(2) & 0.044(3) & --- & 0.01(1) \\ 
QW Ser & 2.65(26) & 17.580(17) & 0.487(66) & 0.081(3) & 0.046(3) & 0.07453 & 0.01(1) \\ 
V650 Peg & 2.52(34) & 18.735(14) & 0.827(114) & 0.222(11) & 0.125(14) & --- & 0.01(1) \\ 
V485 Cen & 3.10(30) & 17.813(38) & 0.672(192) & 0.155(11) & 0.088(14) & 0.04100 & 0.01(2) \\ 
CT Hya & 1.40(26) & 18.547(27) & 0.378(115) & 0.082(4) & 0.046(4) & 0.06520 & 0.01(1) \\ 
\hline
\end{tabular}
}
\end{center}
\end{table*}

\setcounter{table}{2}
\begin{table*}
\caption{List of SU UMa-type DNe identified with $Gaia$ DR2 objects (continued)}
\begin{center}
\scalebox{0.9}{
\begin{tabular}{cccccccc}
\hline
Name & Plx\commenta & $G$\commentb & $G_{\rm col}$\commentc & $A_{G}$\commentd & $A_{\rm BP} - A_{\rm RP}$\commentd & $P_{\rm orb}$\commente & $P_{\rm WZ}$\commenth \\
\hline 
ASASSN-13bm & 2.30(20) & 18.775(20) & 0.783(94) & 0.191(11) & 0.107(16) & --- & 0.01(1) \\ 
ASASSN-18fp & 2.09(28) & 18.288(20) & 0.608(146) & 0.279(18) & 0.157(24) & --- & 0.01(1) \\ 
SDSS J091001.63+164820.0 & 1.87(28) & 18.750(17) & 0.618(68) & 0.081(3) & 0.046(4) & --- & 0.01(1) \\ 
DH Aql & 3.55(19) & 17.966(9) & 0.852(66) & 0.250(17) & 0.141(24) & --- & 0.01(1) \\ 
HO Del & 1.98(22) & 18.211(28) & 0.547(135) & 0.153(14) & 0.086(19) & 0.06266 & 0.01(1) \\ 
GZ Cet & 3.57(23) & 18.365(14) & 0.937(109) & 0.074(2) & 0.042(3) & 0.05534 & 0.01(1) \\ 
AX For & 2.97(18) & 17.798(15) & 0.658(80) & 0.049(1) & 0.028(1) & 0.07850 & 0.01(1) \\ 
V391 Cam & 6.09(5) & 15.532(11) & 0.500(49) & 0.185(2) & 0.104(3) & 0.05620 & 0.01(0) \\ 
CY UMa & 3.18(14) & 17.305(26) & 0.567(136) & 0.031(1) & 0.017(1) & 0.06795 & 0.01(1) \\ 
ASASSN-14ex & 1.42(28) & 19.090(22) & 0.617(123) & 0.104(10) & 0.059(13) & --- & 0.01(1) \\ 
SDSS J162520.29+120308.7 & 2.20(24) & 18.451(12) & 0.732(64) & 0.196(7) & 0.111(9) & 0.09111 & 0.01(0) \\ 
V4140 Sgr & 1.67(16) & 17.719(9) & 0.346(59) & 0.174(5) & 0.098(7) & 0.06143 & 0.01(0) \\ 
DV UMa & 2.60(35) & 18.673(19) & 0.865(96) & 0.017(1) & 0.010(1) & 0.08585 & 0.01(1) \\ 
CRTS J202731.2-224002 & 2.95(16) & 17.890(10) & 0.749(51) & 0.127(5) & 0.072(6) & 0.07000 & 0.01(0) \\ 
V1258 Cen & 2.81(39) & 18.829(43) & 1.004(163) & 0.160(13) & 0.090(18) & 0.08894 & 0.01(1) \\ 
TU Crt & 3.22(14) & 17.754(13) & 0.772(55) & 0.101(3) & 0.057(4) & 0.08209 & 0.01(0) \\ 
NSV 4838 & 1.57(30) & 18.553(19) & 0.538(99) & 0.036(1) & 0.020(1) & 0.06790 & 0.01(1) \\ 
TY PsA & 5.43(7) & 15.795(27) & 0.524(91) & 0.038(1) & 0.021(1) & 0.08410 & 0.01(0) \\ 
AQ Eri & 2.66(12) & 17.211(14) & 0.522(71) & 0.149(5) & 0.084(6) & 0.06094 & 0.01(0) \\ 
V893 Sco & 8.06(5) & 14.647(17) & 0.516(73) & 0.262(4) & 0.148(4) & 0.07596 & 0.01(0) \\ 
V342 Cam & 3.52(11) & 17.647(14) & 0.850(65) & 0.216(7) & 0.122(10) & 0.07531 & 0.01(0) \\ 
AR Pic & 3.14(8) & 17.147(12) & 0.606(57) & 0.109(2) & 0.061(3) & 0.08022 & 0.01(0) \\ 
ASASSN-17ex & 1.54(23) & 18.902(13) & 0.736(77) & 0.351(47) & 0.198(68) & --- & 0.01(0) \\ 
SDSS J100515.38+191107.9 & 2.57(24) & 18.050(15) & 0.728(131) & 0.063(2) & 0.036(2) & 0.07472 & 0.01(1) \\ 
SU UMa & 4.53(3) & 15.292(47) & 0.312(129) & 0.083(2) & 0.046(1) & 0.07638 & 0.00(0) \\ 
SX LMi & 3.08(12) & 16.693(42) & 0.471(160) & 0.071(2) & 0.040(2) & 0.06717 & 0.00(1) \\ 
MASTER OT J042609.34+354144.8 & 5.32(7) & 16.209(17) & 0.743(69) & 0.323(7) & 0.182(9) & 0.06560 & 0.00(0) \\ 
AW Gem & 2.13(29) & 19.081(13) & 0.945(85) & 0.143(14) & 0.080(20) & 0.07621 & 0.00(0) \\ 
SDSS J114628.80+675909.7 & 1.45(16) & 18.203(11) & 0.434(48) & 0.031(0) & 0.017(0) & --- & 0.00(0) \\ 
KV Dra & 2.22(6) & 17.254(18) & 0.440(75) & 0.033(1) & 0.018(0) & 0.05876 & 0.00(0) \\ 
FO And & 1.67(16) & 17.975(51) & 0.499(268) & 0.131(8) & 0.074(8) & 0.07161 & 0.00(1) \\ 
ASASSN-14dw & 1.80(32) & 18.198(9) & 0.626(56) & 0.191(21) & 0.107(30) & --- & 0.00(0) \\ 
LY Hya & 3.14(16) & 17.994(66) & 0.900(296) & 0.106(7) & 0.060(6) & 0.07480 & 0.00(1) \\ 
AK Cnc & 1.69(19) & 18.236(10) & 0.577(73) & 0.076(2) & 0.043(3) & 0.06510 & 0.00(0) \\ 
RX J1715.6+6856 & 1.38(15) & 18.357(19) & 0.494(81) & 0.082(2) & 0.046(2) & 0.06830 & 0.00(0) \\ 
CTCV J1940-4724 & 2.86(6) & 16.732(56) & 0.489(233) & 0.112(5) & 0.063(4) & 0.08090 & 0.00(1) \\ 
CRTS J105122.8+672528 & 0.99(19) & 18.647(22) & 0.373(101) & 0.034(1) & 0.019(0) & 0.05960 & 0.00(0) \\ 
PU CMa & 6.13(3) & 15.245(18) & 0.547(74) & 0.076(1) & 0.043(1) & 0.05669 & 0.00(0) \\ 
DM Lyr & 1.53(12) & 18.427(24) & 0.654(137) & 0.289(19) & 0.163(25) & 0.06541 & 0.00(0) \\ 
CRTS J102842.9-081927 & 1.45(29) & 19.359(14) & 0.845(300) & 0.129(7) & 0.073(6) & 0.03620 & 0.00(1) \\ 
TT Boo & 1.47(20) & 18.955(18) & 0.714(77) & 0.035(1) & 0.019(0) & 0.07594 & 0.00(0) \\ 
ASASSN-16my & 1.61(19) & 18.624(12) & 0.781(72) & 0.412(64) & 0.233(92) & --- & 0.00(0) \\ 
ASASSN-17dg & 2.58(18) & 18.151(16) & 0.899(71) & 0.259(25) & 0.146(36) & --- & 0.00(0) \\ 
CRTS J081418.9-005022 & 1.87(20) & 18.334(20) & 0.705(100) & 0.078(6) & 0.044(8) & 0.07485 & 0.00(0) \\ 
SSS J134850.1-310835 & 2.79(26) & 18.268(25) & 0.949(127) & 0.111(7) & 0.063(9) & --- & 0.00(0) \\ 
1RXS J003828.7+250920 & 2.10(33) & 18.786(25) & 0.914(122) & 0.071(3) & 0.040(5) & --- & 0.00(0) \\ 
1RXS J161659.5+620014 & 2.03(8) & 17.559(36) & 0.541(160) & 0.037(1) & 0.021(1) & --- & 0.00(0) \\ 
WY Tri & 2.04(13) & 17.633(24) & 0.611(101) & 0.177(6) & 0.099(8) & 0.07569 & 0.00(0) \\ 
CRTS J033349.8-282244 & 1.63(22) & 18.081(13) & 0.560(55) & 0.022(0) & 0.012(0) & --- & 0.00(0) \\ 
OT J055717+683226 & 1.48(16) & 18.285(7) & 0.618(33) & 0.206(9) & 0.116(13) & 0.05230 & 0.00(0) \\ 
V1040 Cen & 7.80(3) & 14.035(17) & 0.421(73) & 0.118(2) & 0.066(1) & 0.06049 & 0.00(0) \\ 
SDSS J164248.52+134751.4 & 1.85(12) & 17.826(12) & 0.626(55) & 0.164(3) & 0.092(4) & 0.07710 & 0.00(0) \\ 
SDSS J083845.23+491055.5 & 1.30(23) & 18.329(23) & 0.545(99) & 0.096(2) & 0.054(2) & 0.06920 & 0.00(0) \\ 
ASASSN-13ak & 1.35(25) & 19.042(17) & 0.773(96) & 0.093(3) & 0.052(4) & --- & 0.00(0) \\ 
\hline
\end{tabular}
}
\end{center}
\end{table*}

\setcounter{table}{2}
\begin{table*}
\caption{List of SU UMa-type DNe identified with $Gaia$ DR2 objects (continued)}
\begin{center}
\scalebox{0.9}{
\begin{tabular}{cccccccc}
\hline
Name & Plx\commenta & $G$\commentb & $G_{\rm col}$\commentc & $A_{G}$\commentd & $A_{\rm BP} - A_{\rm RP}$\commentd & $P_{\rm orb}$\commente & $P_{\rm WZ}$\commenth \\
\hline
IR Gem & 3.76(8) & 15.963(32) & 0.538(136) & 0.115(4) & 0.065(5) & 0.06840 & 0.00(0) \\ 
TCP J20100517+1303006 & 1.87(17) & 17.915(21) & 0.733(88) & 0.338(29) & 0.191(41) & --- & 0.00(0) \\ 
CRTS J102705.8-434341 & 1.43(26) & 19.039(61) & 0.853(259) & 0.212(30) & 0.119(41) & 0.07800 & 0.00(0) \\ 
GX Cas & 2.66(23) & 18.264(12) & 1.028(74) & 0.224(26) & 0.127(38) & 0.08902 & 0.00(0) \\ 
NSV 14652 & 1.59(22) & 18.376(17) & 0.813(75) & 0.462(15) & 0.262(20) & 0.07824 & 0.00(0) \\ 
CC Cnc & 1.90(25) & 17.433(21) & 0.550(91) & 0.096(4) & 0.054(5) & 0.07352 & 0.00(0) \\ 
AD Men & 1.83(19) & 18.551(15) & 0.869(69) & 0.161(6) & 0.091(8) & 0.09170 & 0.00(0) \\ 
V660 Her & 1.26(16) & 18.559(15) & 0.667(78) & 0.253(7) & 0.142(9) & 0.07826 & 0.00(0) \\ 
KS UMa & 2.71(11) & 16.821(13) & 0.587(62) & 0.018(0) & 0.010(0) & 0.06796 & 0.00(0) \\ 
MASTER OT J212624.16+253827.2 & 1.88(33) & 18.635(13) & 0.935(60) & 0.218(25) & 0.123(36) & 0.08700 & 0.00(0) \\ 
EI Psc & 6.55(7) & 15.908(5) & 0.904(24) & 0.126(1) & 0.071(2) & 0.04457 & 0.00(0) \\ 
ASASSN-15fv & 0.99(15) & 18.430(26) & 0.469(91) & 0.127(4) & 0.072(5) & --- & 0.00(0) \\ 
BZ Cir & 2.16(7) & 17.063(23) & 0.610(91) & 0.300(13) & 0.169(17) & 0.07900 & 0.00(0) \\ 
KP Cas & 1.79(26) & 19.070(17) & 1.069(77) & 0.283(62) & 0.160(92) & --- & 0.00(0) \\ 
V1006 Cyg & 2.08(7) & 17.287(25) & 0.625(105) & 0.184(6) & 0.104(7) & 0.09903 & 0.00(0) \\ 
SS UMi & 1.89(5) & 16.839(17) & 0.438(68) & 0.079(1) & 0.044(1) & 0.06778 & 0.00(0) \\ 
HS Vir & 2.84(6) & 15.791(16) & 0.418(56) & 0.117(1) & 0.066(1) & 0.07690 & 0.00(0) \\ 
NSV 35 & 4.05(3) & 15.111(29) & 0.430(151) & 0.061(2) & 0.034(1) & 0.07900 & 0.00(0) \\ 
V444 Peg & 2.47(24) & 18.641(27) & 1.132(151) & 0.061(3) & 0.034(4) & 0.09260 & 0.00(0) \\ 
CRTS J214738.4+244554 & 2.01(22) & 18.363(23) & 0.946(105) & 0.130(9) & 0.073(12) & 0.09273 & 0.00(0) \\ 
CRTS J015051.5+332622 & 1.53(16) & 17.787(24) & 0.607(103) & 0.104(3) & 0.059(4) & --- & 0.00(0) \\ 
V632 Cyg & 1.95(7) & 17.196(21) & 0.684(104) & 0.428(19) & 0.242(25) & 0.06377 & 0.00(0) \\ 
QU Aqr & 1.06(13) & 18.159(13) & 0.520(79) & 0.187(3) & 0.105(3) & 0.08330 & 0.00(0) \\ 
V728 CrA & 1.72(11) & 17.045(20) & 0.525(77) & 0.258(16) & 0.145(22) & --- & 0.00(0) \\ 
ASASSN-15aq & 1.61(11) & 17.635(20) & 0.600(113) & 0.038(1) & 0.021(1) & --- & 0.00(0) \\ 
V1239 Her & 3.40(13) & 17.913(25) & 1.150(116) & 0.041(1) & 0.023(2) & 0.10008 & 0.00(0) \\ 
DT Oct & 3.55(4) & 16.223(23) & 0.767(93) & 0.279(6) & 0.157(5) & --- & 0.00(0) \\ 
ASASSN-14eq & 1.54(10) & 17.743(27) & 0.612(113) & 0.022(0) & 0.012(0) & --- & 0.00(0) \\ 
V1227 Her & 1.11(17) & 18.079(14) & 0.557(59) & 0.179(2) & 0.100(2) & 0.06442 & 0.00(0) \\ 
BF Ara & 2.03(23) & 18.212(20) & 1.015(127) & 0.318(40) & 0.180(59) & 0.08417 & 0.00(0) \\ 
CzeV404 & 2.94(5) & 16.040(42) & 0.645(176) & 0.335(13) & 0.189(12) & 0.09802 & 0.00(0) \\ 
1RXS J001538.2+263656 & 1.64(19) & 17.626(23) & 0.679(94) & 0.081(2) & 0.046(2) & 0.10150 & 0.00(0) \\ 
ASASSN-13bc & 0.91(13) & 18.777(39) & 0.655(198) & 0.086(3) & 0.048(2) & --- & 0.00(0) \\ 
ASASSN-15cl & 1.86(27) & 18.931(15) & 1.189(89) & 0.245(50) & 0.139(75) & --- & 0.00(0) \\ 
BR Lup & 1.60(18) & 17.758(10) & 0.796(48) & 0.308(27) & 0.174(39) & 0.07950 & 0.00(0) \\ 
SDSS J225831.18-094931.7 & 3.37(6) & 15.076(95) & 0.445(395) & 0.090(6) & 0.051(4) & 0.08257 & 0.00(0) \\ 
CRTS J155631.0-080440 & 0.97(19) & 18.360(30) & 0.708(152) & 0.424(12) & 0.240(9) & --- & 0.00(0) \\ 
V589 Her & 1.01(17) & 18.884(36) & 0.798(141) & 0.127(3) & 0.071(2) & --- & 0.00(0) \\ 
ASASSN-15rs & 1.77(8) & 16.714(51) & 0.628(207) & 0.441(34) & 0.249(43) & --- & 0.00(0) \\ 
DDE 76 & 1.83(5) & 16.144(30) & 0.392(133) & 0.060(2) & 0.033(1) & --- & 0.00(0) \\ 
VZ Pyx & 3.98(4) & 15.321(32) & 0.707(119) & 0.207(5) & 0.116(5) & 0.07332 & 0.00(0) \\ 
SDSS J173008.38+624754.7 & 1.86(4) & 16.417(31) & 0.534(152) & 0.065(2) & 0.037(1) & 0.07654 & 0.00(0) \\ 
NY Ser & 1.29(5) & 16.075(39) & 0.225(157) & 0.099(3) & 0.056(2) & 0.09750 & 0.00(0) \\ 
KIS J192254.92+430905.4 & 0.74(7) & 17.214(14) & 0.269(66) & 0.270(7) & 0.151(9) & 0.07461 & 0.00(0) \\ 
TU Men & 3.60(6) & 17.081(13) & 1.186(65) & 0.165(3) & 0.093(3) & 0.11720 & 0.00(0) \\ 
NY Her & 0.56(9) & 18.677(73) & 0.520(333) & 0.131(8) & 0.073(5) & --- & 0.00(0) \\ 
WX Hyi & 4.27(3) & 14.628(22) & 0.668(147) & 0.079(2) & 0.045(2) & 0.07481 & 0.00(0) \\ 
YZ Cnc & 4.17(5) & 13.988(87) & 0.486(393) & 0.050(4) & 0.028(2) & 0.08680 & 0.00(0) \\ 
ASASSN-14kf & 2.49(50) & 19.664(5) & 1.799(71) & 0.033(3) & 0.020(5) & --- & 0.00(0) \\ 
QZ Ser & 3.11(9) & 17.245(2) & 1.263(23) & 0.097(1) & 0.055(2) & 0.08316 & 0.00(0) \\ 
CSS J221822.9+344511 & 0.68(12) & 17.670(15) & 0.509(66) & 0.247(4) & 0.139(5) & --- & 0.00(0) \\ 
CRTS J203937.7-042908 & 1.27(15) & 17.859(6) & 1.064(34) & 0.119(3) & 0.067(4) & 0.10572 & 0.00(0) \\ 
V476 Peg & 1.53(6) & 16.901(11) & 1.223(71) & 0.431(13) & 0.245(18) & 0.06370 & 0.00(0) \\ 
\hline
\end{tabular}
}
\end{center}
\end{table*}

\end{document}